\documentclass[aps,twocolumn,prl,showpacs]{revtex4}
\usepackage{epsfig}

\begin{document}

\newcommand \Pomeron {I\!\!P}

\title{Nuclear shadowing and extraction of $F_2^p-F_2^n$  at small $x$ from deuteron collider data}

\author{L. FRANKFURT}
\affiliation{School of Physics and Astronomy, Tel Aviv University, 69978, Tel Aviv, Israel}
\author{V. GUZEY}
\affiliation{Institut f{\"u}r Theoretische Physik II, Ruhr-Universit{\"a}t Bochum,
  D-44780 Bochum, Germany}
\author{M. STRIKMAN}
\affiliation{Department of Physics, the Pennsylvania State University, State
  College, PA 16802, USA}

\preprint{RUB-TP2-06/03}

\begin{abstract}
We demonstrate that leading twist nuclear shadowing leads to
large corrections for  the
 extraction of the neutron structure function $F_2^n$
 from the future deuteron collider data both in the inclusive and 
in the
tagged structure function modes. We suggest several strategies to
 address  
the extraction of $F_2^n$
 and to measure at the same time
the effect of nuclear shadowing via the measurement of the distortion of 
the proton spectator spectrum in the semi-inclusive $e D \to e^{\prime}pX$ process.
\end{abstract}

\maketitle

One of the important components of the electron-ion collider project 
(EIC)~\cite{EIC} is the
measurement of the  nonsinglet structure function
 of the nucleon, $F_2^p-F_2^n$,  
down to rather small values of Bjorken $x$.
A similar 
program is discussed for HERA III.
The main objective of using deuteron beams at HERA is
obtaining more accurate information on the nonsinglet structure function
 of the nucleon at small $x$.
Since $F_2^p$ and $F_2^n$ are rather close at small $x$, nuclear shadowing
significantly affects 
 $F_2^p-F_2^n$,
when extracted from the deuterium data.
Indeed, while 
$R \equiv (F_2^p-F_2^n)/F_2^p \leq 4 \times 10^{-2}$  for 
$x \leq 10^{-2}$, 
 the shadowing correction modifies 
$F_2^{D}$ by 1-2\% and, hence, $R$
is modified
 by twice as much.
Therefore
 it is crucial to determine the value of 
the nuclear shadowing correction,
as well as its uncertainties, 
both to the inclusive structure function $F_2^{D}$ and to the tagged structure function,
when the spectator proton 
(a proton with momentum $\le 0.1$ GeV/c  in the deuteron rest frame) 
 is detected ensuring the kinematics 
maximally close to the scattering off a free nucleon.

In this paper, we  analyse both
of the above mentioned structure functions
using the theory of 
 leading twist nuclear shadowing~\cite{FS99},
which is based on 
the existence of the
deep connection between the phenomena of high energy 
diffraction and  nuclear shadowing demonstrated
within the reggeon calculus for hadron-deuteron scattering
 by Gribov~\cite{Gribov}, 
and the Collins factorization theorem for hard
 diffraction~\cite{Collins}. 
In the case of the tagged
structure functions, we 
 also use the  Abramovski$\breve{{\rm i}}$, 
Gribov,
 Kancheli (AGK) cutting rules~\cite{AGK}, which
 relate shadowing effects for the total and partial cross sections.
We formulate the requirements necessary for 
the extraction of $F_2^n$ from the deuterium data and 
demonstrate that
 a detector with 
 good proton momentum resolution
will be able to determine 
the value of the  shadowing correction.

\begin{figure}[h]
\begin{center}
\epsfig{file=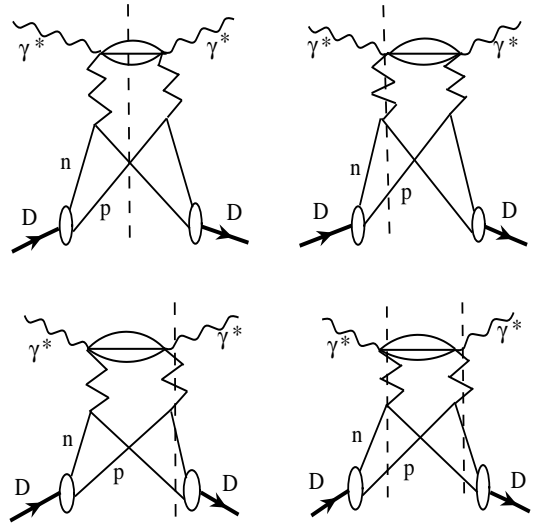,width=9cm,height=8cm}
\vskip -0.25cm
\caption{All possible unitary cuts of the interference diagram giving rise to nuclear shadowing correction to  $\gamma^{\ast} D$ cross section.}
\label{fig:total}
\end{center}
\end{figure}

We begin by discussing
 nuclear shadowing in inclusive DIS off deuterium.
 On the qualitative 
level, the Gribov result for nuclear shadowing  off the deuteron~\cite{Gribov}
 could be understood
as a consequence of 
the interference between the amplitudes for diffractive 
scattering of the projectile  off
the proton and off the neutron of the deuterium target.
Such interference is possible for small $x$, $x \leq 5 \times 10^{-2}$,
where the minimum momentum transfer
to the nucleon, $\sim x m_N$, becomes smaller than the average nucleon
momentum in the deuteron. 
The corresponding double scattering diagram
for the  $\gamma^{\ast}D$ scattering is presented in Fig.~\ref{fig:total}.
The nuclear shadowing correction to the total $\gamma^{\ast}D$ cross section is given by the imaginary part of this diagram, which is obtained by making all possible unitary cuts (denoted by dashed lines in Fig.~\ref{fig:total}). 
The application of the AGK cutting rules
 explicitly demonstrates that the 
imaginary part of the interference graph decreases the total $\gamma^{\ast} D$ cross section by the factor proportional to 
$(1-\eta^2) ({\rm Im} {\cal A})^2$, where ${\cal A}$ is the amplitude for the photon-nucleon diffractive scattering and $\eta={\rm Re}{\cal A} /{\rm Im}{\cal A}$.   

A consideration of the individual unitary cuts 
 in Fig.~\ref{fig:total} demonstrates that the interference diagram decreases the cross section 
of   inelastic interactions
with a single  nucleon
 (top right and bottom left graphs) 
by the factor proportional to  $4({\rm Im}{\cal A})^2$,
and  also results in simultaneous   inelastic
interactions with two  nucleons 
(botton right graph) with
$\sigma_{double}=2({\rm Im} {\cal A})^2$.
 Also, the interference diagram increases the probability of diffraction on the nucleus  by the factor proportional to 
$(1+\eta^2)({\rm Im} {\cal A})^2 $ 
(top left graph),
 see Ref.~\cite{FS96}.

Within the framework of the Gribov theory of nuclear 
shadowing, the deuterium inclusive structure function reads 
\begin{eqnarray}
&&F_2^{D}(x,Q^2)  =  F_{2}^p(x,Q^2)+F_{2}^n(x,Q^2) \nonumber\\
&&-2 \frac{1-\eta^2}{1+\eta^2}\int_{x}^{x_{0}} dx_{\Pomeron} d q_t^2
\, F^{D(4)}_2\left(\beta, Q^2,x_{\Pomeron},t\right) \nonumber\\
&& \times 
\rho_D(4 q_t^2+4 (x_{\Pomeron} m_N)^2) \,,
\label{sh1}
\end{eqnarray}
where $F^{D(4)}$ is the nucleon diffractive structure
function, $\rho_D$ is the deuteron form factor, 
and $|t|=q_t^2+(x_{\Pomeron} m_N)^2$.
Since the $t$-dependence of $\rho_D$ is rather moderate 
(compared to heavier nuclei), the integral in Eq.~(\ref{sh1}) is
sensitive to $F^{D(4)}(t)$ up to $-t \leq 0.05$ GeV$^2$.
 Also, in Eq.~(\ref{sh1}), the factor $(1-\eta^2)/(1+\eta^2)$,
 where 
\begin{equation}
\eta=-\frac{\pi}{2} \frac{\partial \ln(\sqrt{f^D_{i/N}})}{ \partial
 \ln(1/x_{\Pomeron})}=\frac{\pi}{2} (\alpha_{\Pomeron}(t=0)-1) \,,
\label{eta}
\end{equation} 
 accounts for the real part of  the amplitude for
 the diffractive scattering~\cite{AFS99} and leads to a reduction of
the shadowing by about  20\%. 
Note that in the reggeon calculus derivation~\cite{Gribov}, it was assumed that $\eta=0$,
which is natural for the amplitudes slowly increasing with energy.
This is not the case for DIS and, hence, $\eta$ should be taken into account.

\begin{figure}[ht]
\begin{center}
\epsfig{file=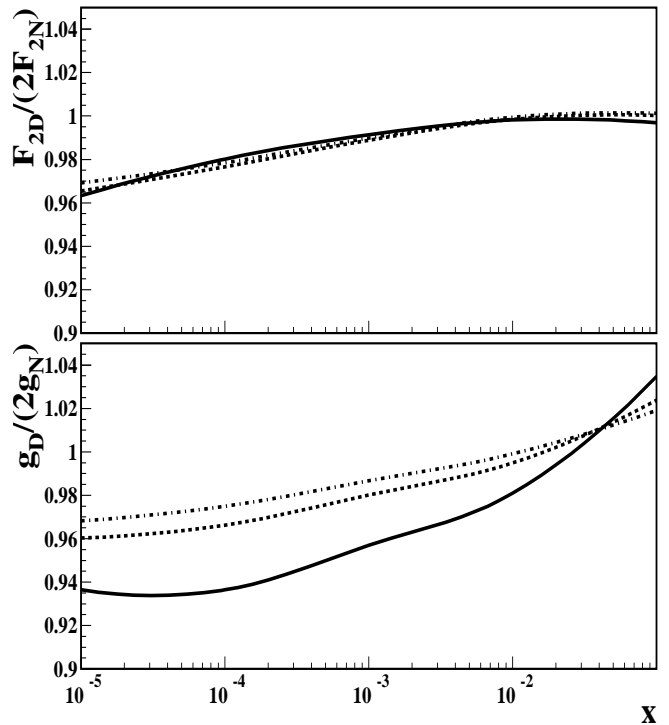,width=9cm,height=11cm}
\vskip -0.25cm
\caption{The ratios $F_2^D/(F_2^p+F_2^n)$ and $g_D/(2g_N)$ as functions of $x$. The solid curve corresponds to $Q=2$ GeV; the dashed curve corresponds to $Q=5$ GeV; the  dash-dotted curve corresponds to $Q=10$ GeV.}
\label{fig:d1}
\end{center}
\end{figure}

\begin{figure}[t]
\begin{center}
\epsfig{file=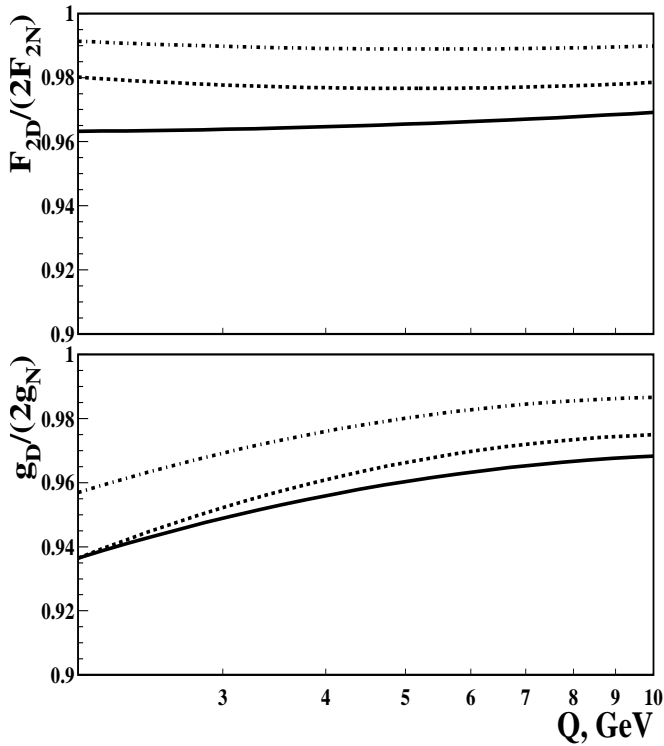,width=9cm,height=11cm}
\vskip -0.25cm
\caption{The ratios $F_2^D/(F_2^p+F_2^n)$ and $g_D/(2g_N)$ as functions of $Q$. The solid curve corresponds to $x=10^{-5}$; the dashed curve corresponds to $x=10^{-4}$; the dash-dotted curve corresponds to $x=10^{-3}$.}
\label{fig:d2}
\end{center}
\end{figure}

Using the QCD factorization theorem for 
hard diffraction~\cite{Collins},
it is possible to extend the 
Gribov theory in order to calculate  nuclear shadowing 
for the quark and gluon parton densities of the deuteron, $f_{j/D}$, at small $x$ as follows, see Ref.~\cite{FS99} and subsequent publications~\cite{FGMS02,FGS03}, 
\begin{eqnarray}
&&f_{j/D}(x,Q^2) =  f_{j/p}(x,Q^2)+f_{j/n}(x,Q^2) \nonumber\\
&&-2 \frac{1-\eta^2}{1+\eta^2}
\int_{x}^{x_{0}} dx_{\Pomeron} d  q_t^2 \, f^{D}_{j/N}\left(\beta, Q^2,x_{\Pomeron},t\right) \nonumber\\
&&\times \rho_D(4 q_t^2+4 (x_{\Pomeron} m_N)^2) \,.
\label{shdeu}
\end{eqnarray}
The results of the calculation of the ratios $F_2^D/(F_2^p+F_2^n)$ and $g_D/(2g_N)$   
using the H1 diffractive fit~\cite{H1:1994} for $F^{D(3)}$
are presented in Figs.~\ref{fig:d1} and \ref{fig:d2} for a range of
$x$ and $Q$. In this calculation we used the Paris 
NN potential. One can see that
 substantial  shadowing is expected for the small $x$ region.

Note also that the nuclear shadowing correction to $F_2^D$ for $Q^2 \leq 1$ GeV$^2$ 
is expected to be  substantially larger than that for $Q^2 \sim 4$ GeV$^2$. The
enhancement of diffraction at small $Q^2$ by 
higher twist effects such as, for instance,  vector meson production,
will increase nuclear shadowing up to a factor of two.
Note that the application of the Gribov formalism to nuclear shadowing
 in the NMC kinematics ($Q^2 \le 2$ GeV$^2$),
where the double scattering term dominates, leads to a very good 
description of the data, see e.g.~\cite{Kaidalov}.

Since the diffractive cross sections are likely to be known with accuracy of
about 
10\% for small $t$, 
 it appears that the accuracy of the calculation of the nuclear shadowing
 correction to
 the inclusive cross section will be better than 0.2\%
 for a wide range of $Q^2$. Correspondingly, the theoretical uncertainty
for the ratio of $F_2^n/F_2^p$ will not exceed 0.4\%, which is likely
to be smaller than possible experimental
systematic errors.

A   strategy, which is complimentary to 
the inclusive measurement of $F_2^{D}$,
 is the use the neutron and proton tagging.
We will primarily focus on tagging of scattering off a neutron via
the detection of the spectator protons and will
only
briefly comment on other possibilities.

It was pointed out above that according
 to the AGK cutting rules~\cite{AGK}, the overall 
correction of the nuclear shadowing effect 
 for nondiffrative  events  is four times larger than for the inclusive 
scattering. However, it is concentrated at relatively large transverse
 momenta ($k_t$) of the protons and depends on $k_t$ rather strongly.
Hence, two strategies will be possible. One would be to select only very low 
$p_t$ protons with a gross loss of statistics. The other, more
 promising
approach  is to measure the $p_t$ dependence of the
 spectrum up to $p_t \sim 250$ MeV/c
and then to use this for the verification
 of the theory and for the measurement of 
the effects 
of the rescattering (nuclear shadowing) in an independent way
with the subsequent 
correction of the data for this effect.
 Provided good momentum resolution of the  proton
 spectrometer, one would be able to make longitudinal momentum 
cuts to suppress/increase  the shadowing effect (the shadowing
 effects are minimal for $|2p_N/p_{D}-1|\leq 0.1$).

The following analysis demonstrates the dependence of the nuclear shadowing correction to the $\gamma^{\ast} D \to  p X$ cross section on the transverse momentum of the spectator proton $p_t$.
The impulse approximation expression for the differential
 cross section of interest reads
\begin{equation}
\frac{d \sigma^{\gamma^{\ast} D \to  p X}}{d^3 p}\Big|_{{\rm IA}}=\sigma^{\gamma^{\ast} n} 
(2-2x_L) (u^2(p)+w^2(p)) \,,
\label{ia}
\end{equation}
where $x_L=E_p/E_{D} \approx (1- p_z/m_N)/2$ is the Feynman $x$ of the spectator proton;
$p=(p_t,p_z)$ is the three-momentum of the detected (spectator) proton 
in the deuteron rest frame; $u$ and $w$ are the $S$-wave and $D$-wave components of the deuteron momentum space wave function normalized as
$4 \pi \int dp\, p^2 (u(p)^2+w(p)^2)=1$; the factor $2-2x_L$ is
the M{\"u}ller flux factor.
  The presence of 
leading twist nuclear shadowing adds a nuclear shadowing correction
 to Eq.~(\ref{ia}), so we find for the  complete cross section 
\begin{eqnarray}
&&\frac{d \sigma^{\gamma^{\ast} D \to  p X}}{d^3 p}=\sigma^{\gamma^{\ast} n} (2-2x_L)(u^2(p)+w^2(p)) -
 \frac{3-\eta^2}{1+\eta^2}\nonumber\\
&&\times  \int_{x}^{x_{0}} dx_{\Pomeron} \int \frac{d^2 q_t}{\pi}\, F^{D(4)}_2\left(\beta, Q^2,x_{\Pomeron},t\right) \Bigg[u(p) u(p^{\prime}) \nonumber\\
&&+w(p)w(p^{\prime})(\frac{3}{2}\frac{(p \cdot p^{\prime})^2}{p^2 p^{\prime 2}}-\frac{1}{2}) \Bigg] \,,
\label{spec:full}
\end{eqnarray}
where $p^{\prime}=p+q_t+(x_{\Pomeron} m_N) e_z$.
 The additional factor of $3-\eta^2$ in front of the shadowing correction
 is 
a reflection of the AGK cutting rules. In the 
considered case, the shadowing correction is given by the sum of the two top graphs  in Fig.~\ref{fig:total}. The left graph corresponds to the enhancement of
 the diffractive processes by the factor of $(1-\eta^2)({\rm Im}{\cal A})^2 $, while  
 the right graph corresponds to screening  from the single inelastic subprocesses and is proportional to $-4({\rm Im}{\cal A})^2$.
In Eq.~(\ref{spec:full}), the factor $u(p) u(p^{\prime})+w(p)w(p^{\prime})(3(p \cdot p^{\prime})^2/(2 p^2 p^{\prime 2})-1/2)$ is simply the unpolarized deuteron density matrix. 
Also,
we neglected production
of deuterons in the diffractive channel, which would  somewhat increase the 
shadowing effect.

\begin{figure}[t]
\begin{center}
\epsfig{file=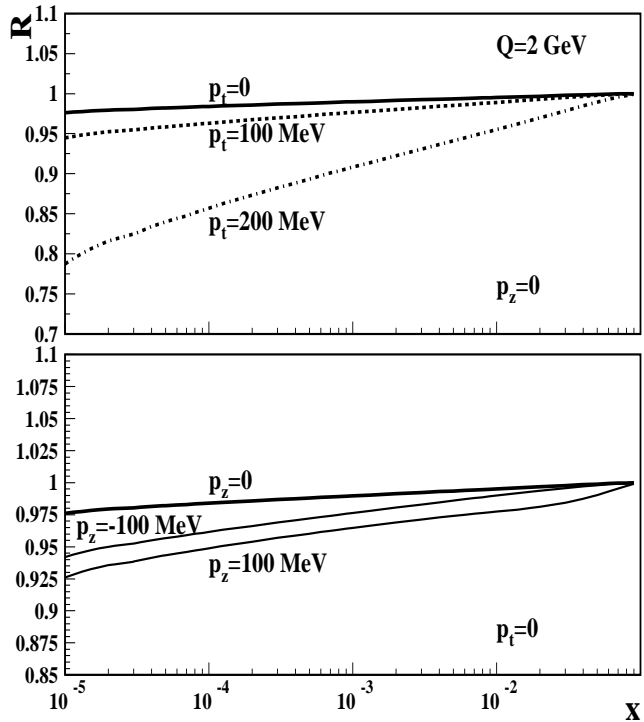,width=9cm,height=11cm}
\vskip -0cm
\caption{The suppression of the proton spectrum by the nuclear shadowing correction. Top panel: The solid curve corresponds to $p_t=0$; the  dashed curve corresponds to $p_t=100$ MeV; the dash-dotted curve corresponds to  $p_t=200$ MeV. Bottom panel: $p_t=0$ and $p_z=(0, -100, +100)$ MeV/c.}
\label{fig:spec}
\end{center}
\end{figure}

As one can see from Eq.~(\ref{spec:full}), nuclear shadowing suppresses the spectrum of the produced protons. This effect can be quantified by considering the ratio $R$ of the complete expression given by Eq.~(\ref{spec:full}) to the impulse approximation expression given by Eq.~(\ref{ia}). Figure~\ref{fig:spec} depicts the ratio $R$ as a function of Bjorken $x$ for $p_t=(0, 100, 200)$ MeV/c and $p_z=0$ 
(top panel) and for $p_t=0$ and $|p_z|=100$ MeV/c (bottom panel).
The calculation is made at $Q=2$ GeV.

Two features of Fig.~\ref{fig:spec} are of interest and importance. Firstly,
nuclear shadowing works to decrease the ratio $R$ as $p_t$ increases. 
This is expected from the general picture of nuclear shadowing since large $p_t$ corresponds to small transverse distances between the two nucleons, which leads to their shadowing, as was first pointed out by Glauber~\cite{Glauber}
and later generalized by Gribov~\cite{Gribov}.
Secondly, the suppression of $R$ at large $p_t$ is strikingly large.
This is a common feature of semi-exclusive reactions with nuclei. Indeed, at large $p_t$, while the impulse approximation term is suppressed by the nuclear wave function, the rescattering term survives and gives the dominant contribution. An example of this effect in electron-deuteron interactions in the TJNAF kinematics can be found in Ref.~\cite{FGMSS}.

Note also that there is a non-spectator contribution to the nucleon spectrum, 
which originates predominantly
from diffractive scattering off the proton and  dominates at large
$p_t$, $p_t \geq 300$ MeV/c.
As a result of the AGK cancellations, the non-spectator
 contribution is given by the impulse approximation. The contribution has a broad $p_t$ distribution, $\propto e^{-B p_t^2}$ with $B \sim 7$ GeV$^{-2}$, and 
can be subtracted using the measurements at, for instance,
 $p_t \geq 400$ MeV/c.
It  appears that it will be possible to use 
 most of the spectator protons from most of the
deuteron wave function
in the tagged method.
Hence the spectator tagging
 will not lead to a loss of
statistics.
The method will allow to introduce corrections for nuclear shadowing
for small $p_t$ with high precision,
which would lead to theoretical errors in
the determination of $F_2^n$ 
at  the level of a fraction of percent.

One can also use simultaneous tagging of protons and neutrons, when 
both neutron and proton  are detected in the reactions $\gamma^{\ast} D \to n X$ and $\gamma^{\ast} D \to p X$.
In this case, 
nuclear shadowing will cancel in the ratio $\sigma^{\gamma^{\ast} D \to n X}/\sigma^{\gamma^{\ast} D \to p X}$
and the main errors in the measurement of $F_2^n$ will be due to the
determination of relative efficiencies of the proton and  neutron taggers.

One could also try to obtain the ratio $F_2^n /F_2^p$ from the comparison of the rate of the tagged proton scattering events 
with the neutron spectator 
to inclusive $e\,D$ scattering.  Such a strategy could also have certain merits
as it avoids the issue of luminosity and does not require a leading proton spectrometer. The disadvantage of this strategy 
is the sensitivity to the nuclear shadowing effects and errors in the acceptance of
the neutron detector.  One possible way to deal with the latter problem will be to perform 
measurements at very small $x$ and large
energies,  where the
$ep$ and $en$ cross sections
are equal 
to better than a fraction of 1\% and, hence, one would be able to cross
check acceptances of the proton and neutron detectors.

Note also that taking proton data from an independent run will 
potentially lead to another
set of issues such as relative luminosity, the use of different
 beam energies, etc., which is likely to be on the level of 1\%.

In our analysis we neglect possible non-nucleonic components of the
deuteron wave function such as kneaded (six-quark) components, etc.
Current estimates put an upper limit on the probability of
such components on the
level of less than one percent, and they are expected to modify predominantly
the high-momentum
component of the deuteron wave function. Hence, the non-nucleonic 
components should give a
very small (less than one percent) correction for the spectator momenta  $\le 200$ MeV/c.
 Moreover, the experiments at
the  EIC
looking for production of  baryons such as $\Delta$ and 
$N^{\ast}$ in the spectator 
kinematics, $x_L(\Delta,N^{\ast})\ge 1/2$, would allow to put a more stringent
upper  limit on (discover) non-nucleonic components of the deuteron wave
function.

In conclusion, we have demonstrated 
that  a combined analysis of 
inclusive and
 semi-inclusive scattering off the deuteron 
coupled  with 
a high resolution proton spectrometer 
will allow for the
  measurement of $F_2^n$ at small $x$ with the theoretical
uncertainty of better than 0.5\%.
 It would be a challenge to reduce
the experimental systematic errors to a comparable level.
The measurement of the shape of the spectator spectrum would allow to
determine
 nuclear shadowing in the deuteron with precision by far
exceeding that possible 
in  the inclusive measurements.

This work was supported by GIF, Sofia Kovalevskaya 
Program of the AvH Foundation and DOE. We thank M.~Klein  for questions,
which sped up our plans to perform this analysis, and for insightful
comments.

\end{document}